\documentclass[a4paper]{mem}
\usepackage{natbib}
\usepackage{graphicx}
\usepackage[a4paper]{hyperref}
\idline{00}{0}
\begin{document}
   \title{Clusters of galaxies in the mid-infrared
\thanks{Based on observations with ISO, an ESA project with instruments
funded by ESA Member Ststes (especially the PI countries: France,
Germany, the Netherlands and the United Kingdom) and with participation
of ISAS and NASA}
}

   \author{A. Biviano \inst{1}, 
   L. Metcalfe \inst{2,3},
   D. Coia \inst{4},
   B. McBreen \inst{4},
   B. Altieri \inst{2}, \\
   R. Perez-Martinez \inst{2,3},
          \and
   C. Sanchez-Fernandez \inst{2}\fnmsep
}

 \offprints{A. Biviano}

\institute{INAF/Osservatorio Astronomico di Trieste, via G.B. Tiepolo 11, 
34131, Trieste, Italy, \email{biviano@ts.astro.it}
\and XMM-Newton Science Operations Centre, European Space Agency,
Villafranca del Castillo, P.O. Box 50727, 28080 Madrid, Spain
\and ISO Data Centre, European Space Agency, Villafranca del Castillo, 
P.O. Box 50727, 28080 Madrid, Spain
\and Physics Department, University College Dublin, Stillorgan Road, 
Dublin 4, Ireland
}

\abstract{We describe the results of observations of galaxy
clusters conducted with ISOCAM on-board the {\em Infrared Space
Observatory.} Our research is aimed at understanding the
processes driving galaxy evolution in dense environments, free from
the bias of dust extinction. The results reveal quite a complex
picture: the star-formation activity of cluster galaxies does not show
a simple evolution with redshift, but also depends on
the dynamical status and evolutionary history of the clusters.
 }
\authorrunning{A. Biviano et al.}  \titlerunning{Clusters of galaxies in
the mid-infrared} \maketitle

\section{Introduction}

\begin{figure*} \centering \resizebox{\hsize}{!}{
\includegraphics[clip=true]{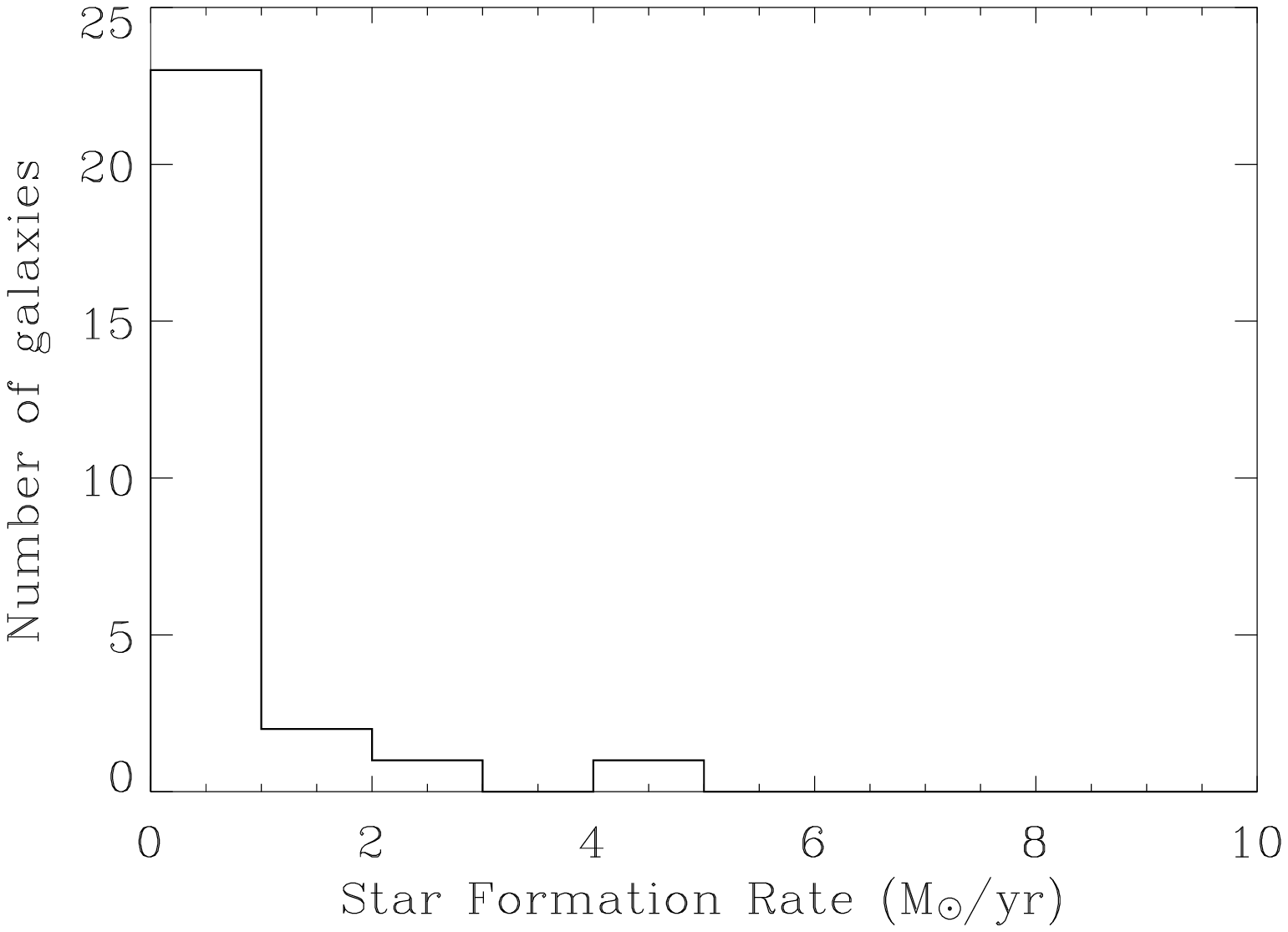}
\includegraphics[clip=true]{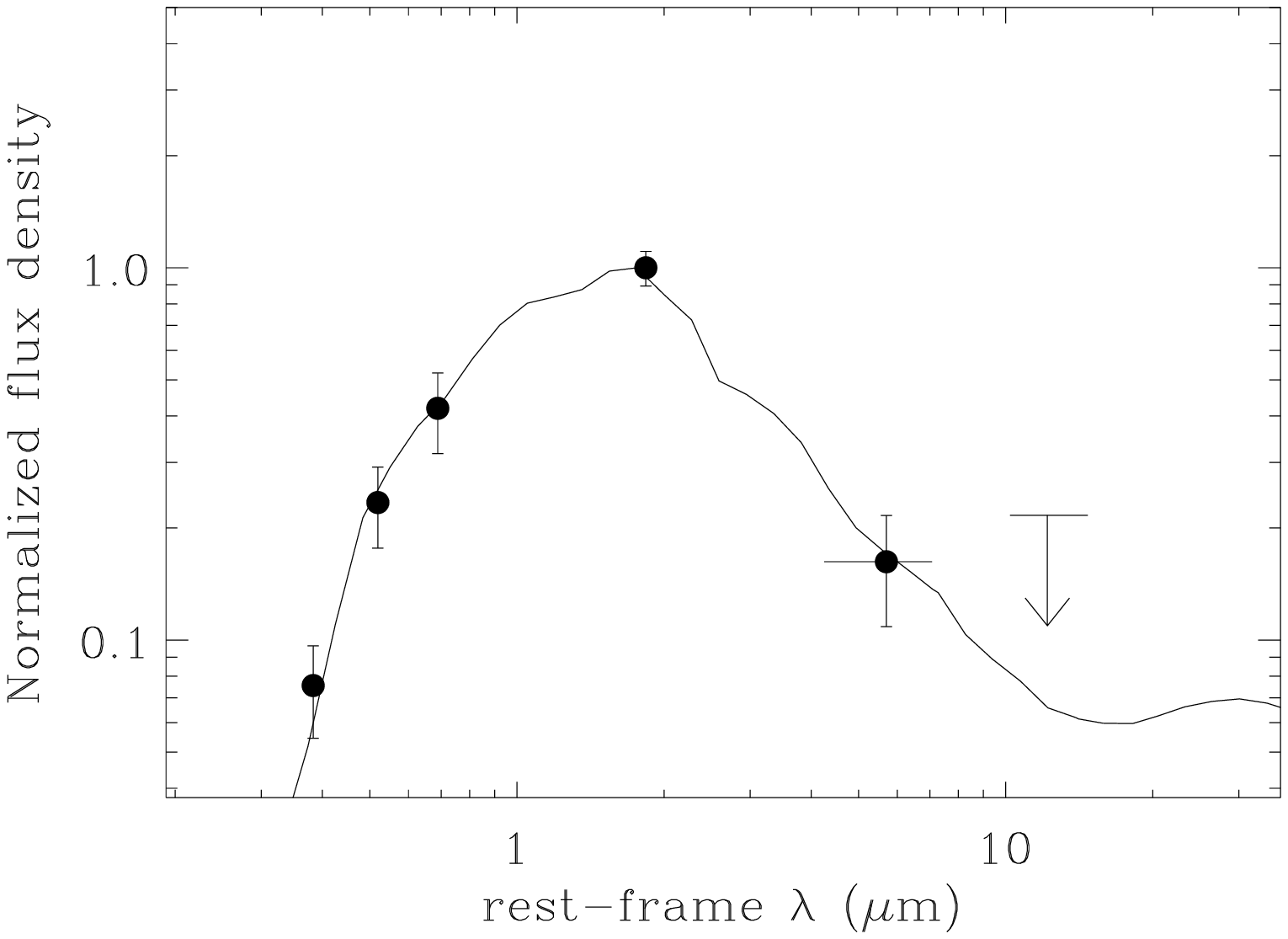}} 
\caption{Left-hand panel:
histogram of the SFRs of the A2218 cluster galaxies detected with
ISOCAM. Right-hand panel: The average spectral energy distribution
of the ISOCAM-detected A2218 cluster galaxies.  The solid line
indicates the best-fit GRASIL model (a passive elliptical).}
\end{figure*}

Many lines of evidence \citep[e.g.][]{van96,lew02} indicate that the
stellar populations of galaxies in cluster cores are old, with most of
their stars formed at high redshifts.  There is nonetheless
observational evidence suggesting that at least a fraction of the
cluster galaxies were more active in the past, namely: (a) the
fraction of blue cluster galaxies increases with redshift -- the
so-called 'Butcher-Oemler' (BO, hereafter) effect \citep{but78,but84},
and (b) the fraction of early-type clusters decreases with redshift
\citep{dre97,fas00}. Presumably, both the colour and the morphological
evolution of the cluster galaxy population derive from the
transformation of blue spirals into S0's, as they fall into clusters
from the field \citep{kat03}. Several physical mechanisms could
operate this transformation \citep[e.g.][]{oka03}, all resulting in
the galaxy gas depletion, and a decrease of the star formation (SF)
activity. Some of these processes induce a starburst phase before the
gas depletion, and some do not.

In order to obtain an unbiased picture of the evolutionary history of
cluster galaxies, it is important to observe them in
the infrared (IR). In fact, dust can mask ongoing SF in a galaxy, and
the amount of extinction increases with the amount of SF
\citep[e.g.][]{sil98}. In this paper we present the results of the
mid-IR (MIR) observations of three galaxy clusters.

\section{The data-set}

\begin{figure*}
\centering
\resizebox{\hsize}{!}{
\includegraphics[clip=true]{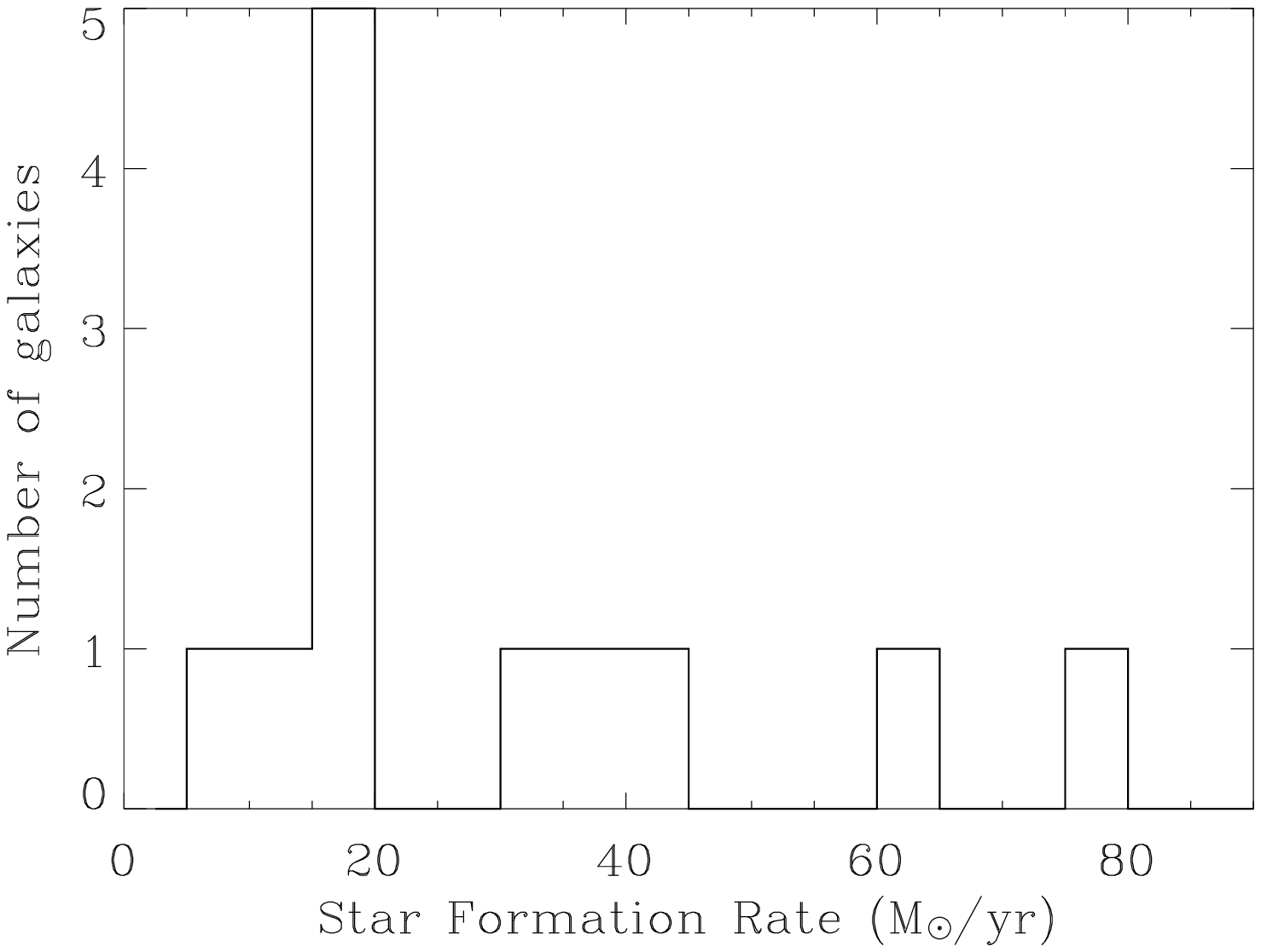}
\includegraphics[clip=true]{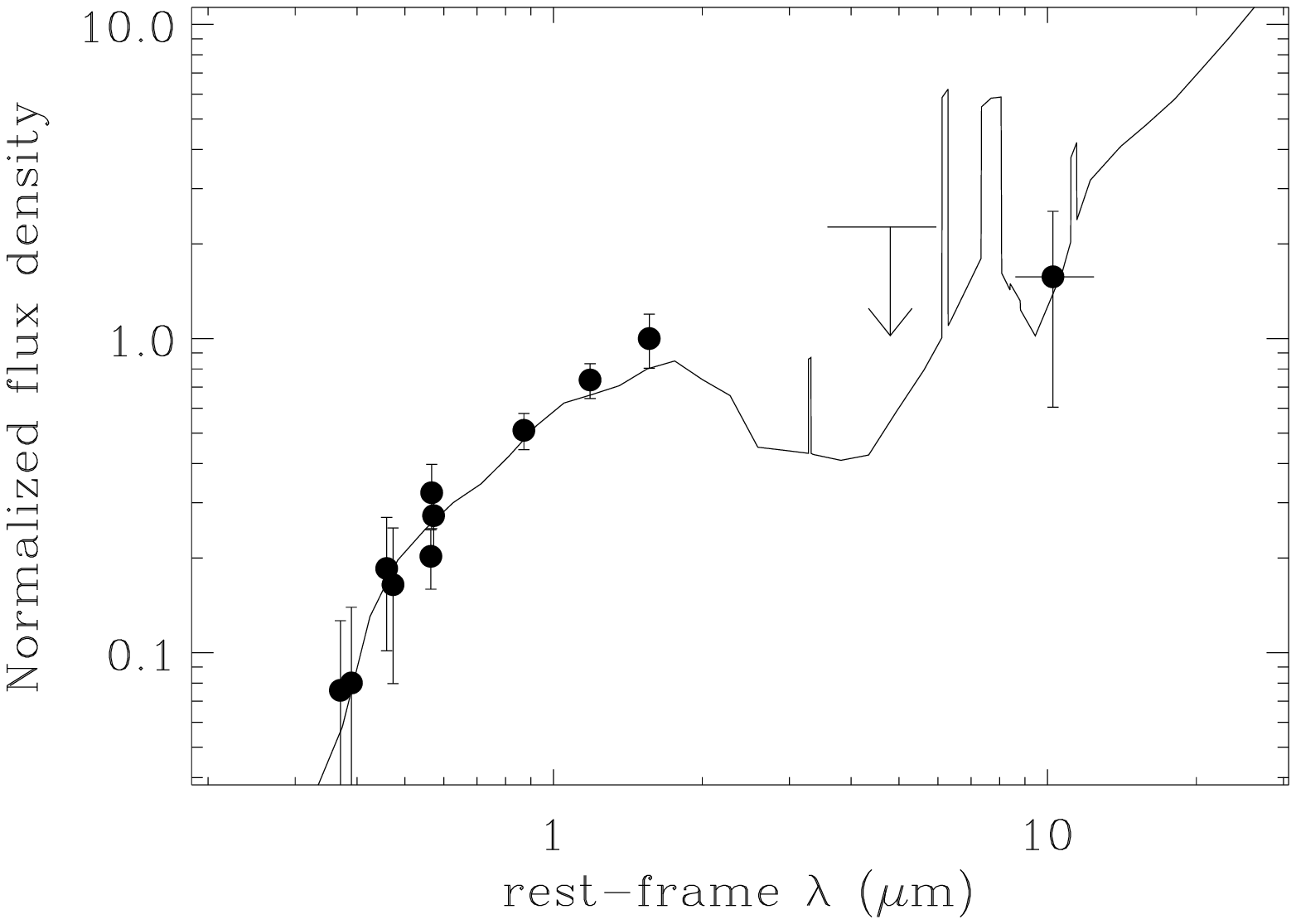}}
\caption{Left-hand panel: histogram of the SFRs
of the Cl0024+1654 cluster galaxies detected with ISOCAM.
Right-hand panel: The average spectral energy distribution of
the ISOCAM-detected Cl0024+1654 cluster galaxies.
The solid line indicates the best-fit GRASIL model (an aged starburst).}
\end{figure*}

The MIR data used in this paper
come partly from a survey -- described in detail in
\citet{met03} -- conducted with ISOCAM onboard the {\em Infrared Space
Observatory} (ISO) \citep{ces96}, and partly from the ISO archive. A
detailed description of the data reduction, source detection and
photometry, is given in \citet{met03}. The three clusters considered
here are A2218, A2219, and Cl0024+1654. A2218 and Cl0024+1654 have
been imaged in two (wide) ISOCAM filters, LW2 and LW3, centered at 6.7
and 14.3 $\mu$m, respectively; only the LW3 filter has been used for
the A2219 field.

These three clusters span a redshift range 0.175--0.395, and the
physical areas covered by observations vary from $0.8 \times 0.8$
Mpc$^2$ (A2218 and A2219) to $2 \times 2$ Mpc$^2$
(Cl0024+1654)\footnote{In this paper we use $H_0=70$
km~s$^{-1}$~Mpc$^{-1}$, $\Omega_0=0.3$ and $\Omega_{\Lambda}=0.7$.}.
The final data-set contains MIR detections of 43 cluster members.  The
limiting sensitivities reached were 34 and 95 $\mu$Jy at,
respectively, 6.7 and 14.3 $\mu$m, in the A2218 field, and 468 and 160
$\mu$Jy at 14.3 $\mu$m in, respectively, the A2219, and Cl0024+1654
field.

The cluster membership is established by cross-correlation of the MIR
sources with catalogues of optical and near-IR sources from the
literature. The optical and near-IR data were also used to define
the spectral energy distributions (SEDs, hereafter) of the ISOCAM
sources, which were then fit with a battery of GRASIL models
\citep{sil98}. The best-fit models were then used to compute the
galaxy total IR luminosities, from which the SF rates (SFRs hereafter)
were derived using the relation of \citet{ken98}.

\section{Results}

{\bf A2218} is a massive cluster of galaxies (l.o.s. velocity
dispersion $\sigma_v \sim 1400$ km~s$^{-1}$) at a mean redshift
$z=0.175$. We detected 27 MIR cluster members, most of them at 6.7
$\mu$m only. The left-hand panel of Figure~1 shows the distribution of
SFRs for the 27 A2218 cluster sources; most of them have negligible
SFRs. As a matter of fact, the SEDs of most MIR-selected cluster
members decline from the near-IR to the MIR, and are best fit by
GRASIL models of aged, passively evolving ellipticals. This is best
seen in the average SED for the 27 cluster sources, shown in the
right-hand panel of Figure~1, built by normalizing each source SED
with its $H$-band flux density. The solid line is the best-fit GRASIL
model, an aged, passively-evolving elliptical.  The MIR emission of
most A2218 cluster sources is the Rayleigh-Jeans tail of the
photospheric emission from cold stars \citep[see also][ for galaxies
in the nearby Virgo and Coma clusters]{bos97}.

Among the 27 mid-IR cluster sources, 7, however, do show evidence of
some SF activity. These are among the faintest ISOCAM-detected cluster
members in the optical and near-IR. Hence, even if their SFRs are
small ($<5$~$M_{\odot}$~yr$^{-1}$), their IR luminosities are
comparable or even higher than their near-IR and optical luminosities.
SF activity therefore seems to be still ongoing in the less massive
cluster members \citep[reminiscent of the 'downsizing' effect
described in, e.g.,][]{pog04}.

{\bf A2219} is a massive cluster of galaxies with a velocity dispersion very
similar to that of A2218, and a slightly higher mean redshift
($z=0.225$). Three cluster members were detected; they all display
significant SF activity, with SFRs between 10 and 24
$M_{\odot}$~yr$^{-1}$. Their SEDs are best fit by models of massive
spiral galaxies, or of aged starbursts. At variance with what we saw
in A2218, here the strong MIR emission of the A2219 cluster galaxies
implies they contain a significant amount of dust that re-emits an
intense stellar radiation field in the IR.

{\bf Cl0024+1654} is a cluster at $z=0.395$, with a velocity
dispersion of $\sim 900$ km~s$^{-1}$.  In total, 13 cluster sources
were detected at 14.3 $\mu$m. They all show a substantial amount of SF
activity, as can be seen from their SFR distribution shown in the
left-hand panel of Figure~2. Their SEDs are best fit by models of
massive (young) spirals, or aged starbursts. This is best seen from the
right-hand panel of Figure~2, where we show the average SED of the 13
MIR cluster sources, with the best-fitting model (an aged starburst
galaxy) overlaid.  Note the very different shape of this SED and of
the average SED of A2218 cluster members (Figure~1). As in A2219,
also in Cl0024+1654 the IR emission of MIR-selected cluster galaxies
probably arises from an intense stellar radiation field absorbed and
re-emitted by dust.

Interestingly, the spatial distribution of the MIR cluster members
in Cl0024+1654 is significantly different from that of the
optically-selected cluster members, the former being less centrally
concentrated. This suggests that the MIR cluster galaxies are a
distinct population from the other members of this cluster.

\section{Conclusions}
We detected 43 MIR emitting galaxies, members of three clusters.  In
one of them, A2218, just as in several nearby clusters \citep[see
also][]{bos97}, most mid-IR-selected galaxies have negligible or small
SFRs, and the IR emission is the Rayleigh-Jeans tail of cold stellar
photospheres. However, in both A2219 and Cl0024+1654, the detected MIR
sources have high SFRs, several tens $M_{\odot}$~yr$^{-1}$, and their
emission is likely to come mostly from dust re-processed stellar
radiation (although we cannot exclude an AGN contribution in some of
them). Comparing their SFRs with those derived from optical lines, it
is found that up to 90\% of their SF activity is hidden by dust.

At first, it might seem that we are seeing an IR analogue of the
BO effect, since the average SFR of MIR
cluster galaxies seems to increase with the mean redshift of the
cluster. However, not all MIR star-forming
cluster galaxies are part of the BO population, and a
fraction of them are actually found along the main colour-magnitude
sequence of passive early-type galaxies. Their red colour is however
not an indication of an aged stellar population, but of significant
dust extinction.

It is important to understand why there is such a spread of the
average SFR of cluster galaxies among different clusters. Part of the
difference likely arises because of the different surveyed areas.
Since the MIR cluster galaxies with high SFRs seem to avoid the inner
cluster regions, we could have missed this population in, e.g.,
A2218. However, the surveyed area of A2219 is comparable to that of
A2218, and yet the MIR populations of the two clusters have different
SFR-distributions. Probably, cluster properties other than their
redshifts play an important r\^ole in determining the
average SFR of cluster galaxies. In fact, A2218 and Cl0024+1654 are
located at the same redshifts as, respectively, A1689 and A370, yet
their member galaxies have very different SFR distributions
\citep{duc02,coi04}. \citet{bek01} suggested that 
the disturbed gravitational tidal field induced
by an infalling sub-cluster might trigger SF
in cluster galaxies. However, all three clusters
considered here show evidence of substructures. It seems likely that the SF
properties of cluster galaxies are determined not only by the
dynamical status, but also by the dynamical history of their cluster
\citep[see][ for a thorough discussion]{biv04}.


\bibliographystyle{aa}

\end{document}